# Noninvasive identification of carbon-based black pigments with pump-probe microscopy

Heidi V. Kastenholz[1], Michael I. Topper[2,†], Warren S. Warren[1,2,3,4], Martin C. Fischer[1,2], and David Grass[1]

1. Department of Chemistry, Duke University, Durham, NC 27708, USA.
2. Department of Physics, Duke University, Durham, NC 27708, USA.
3. Department of Biomedical Engineering, Duke University, Durham, NC 27708, USA.
4. Department of Radiology, Duke University, Durham, NC 27710, USA.

† Presently at Department of Physics, University of Colorado, Boulder, CO 80309, USA.

**Abstract**: Carbon-based black pigments, a widely used class of pigments, are difficult to differentiate with the noninvasive techniques currently used in cultural heritage science. We utilize pump-probe microscopy to distinguish four common carbon-based black pigments as pure pigments, as two-component black pigment mixtures, and as a mixture of a black and a colorful pigment. This work also demonstrates that even nominally "homogeneous" pigments present remarkable, and useful, heterogeneity in pump-probe microscopy.

**Introduction**
There is an unmet need in cultural heritage science for non-invasive identification of carbon-based black pigments, which are broadly used in paintings, drawings, and prints either by themselves or for shading another pigment (*1*). These pigments are easily produced through controlled burning of a material such as wood, bone, or oil (resulting in charcoal, bone black, and lamp black) or they occur naturally, such as graphite (*2*). They have been identified in some of the oldest pieces of art known to date, such as the cave paintings of Nawarla Gabarnmang in northern Australia (*3*). As their sourcing and cost are not prohibitive, carbon-based black pigments still represent one of the primary black pigment sources. There are two classification schemes for carbon-based black pigments (*1, 2, 4-6*). The first is by the level of order in the carbon network of the material, but this information is often inaccessible for black pigments incorporated into artwork. The second classification is based on the materials' origin, but reliable information for pigments in historic works, particularly carbon-based based pigments where the naming conventions are tangled, is often missing.

Material identification is essential for conservation of a work of art and provides insight into its historical context and provenance. In that respect, carbon-based blacks are problematic. The most specific method for identification of carbon-based black pigments, scanning electron microscopy with energy dispersive spectroscopy (SEM-EDX), has been used to distinguish carbon-based black pigments by morphology and, occasionally, elemental composition in pure reference samples (*1, 2, 5, 7, 8*). However, this requires physical removal of a cross-section from the work. Another approach is thermogravimetric analysis and differential scanning calorimetry. This study allowed for characterization of pure reference samples, but also requires invasive sampling (*9*). The go-to non-invasive methods in cultural heritage science are linear reflectance techniques, such as fiber-optic reflectance spectroscopy, hyperspectral imaging, multispectral imaging, and Raman spectroscopy due to their ease of use and portability (*10-15*). Unfortunately, linear reflectance measurements of carbon-based black pigments are featureless in the visible-NIR region (*11, 12*), and Raman spectroscopy is not well suited as the dominant spectral features of all carbon-based black pigments are the same. Presence of a carbon-based black pigment is confirmed by two characteristic peaks at approximately 1580 cm$^{-1}$ and 1350 cm$^{-1}$ (*6, 16-19*). The 1580 cm$^{-1}$ peak (G Band) is the characteristic Raman peak for crystalline graphite (*6, 16*). The 1350 cm$^{-1}$ peak (D or Disorder Band) is used as a measure of disorder in the carbonaceous material; it suggests the presence of heteroatoms in the graphitic structure, in-plane defects, or defects at the edge of the aromatic structure such as a tetrahedral carbon rather than the expected trigonal planar carbon (*6, 16*). Raman studies have attempted to delineate reference pigments by using the minute differences between these peaks, but so far, the results have not been encouraging (*6, 17-19*).

Features in Fourier transform infrared (FTIR) spectra can be used to distinguish between reference carbon-based black pigments (*5, 20*). However, FTIR spectra from paintings have had mixed results. The spectra are either dominated by signals from the ground layer and the resin varnish; any features that would indicate a carbon-based pigment is overpowered by the other materials present (*21*) or they rely on other compounds present, like hydroxyapatite in ivory and bone black (*22*). X-ray fluorescence (XRF) is another noninvasive technique used in cultural heritage science; it cannot distinguish carbonaceous materials (*23*) but reveals secondary elements like Ca and P in compounds like hydroxyapatite to support an identification or rule them out (*24-27*). X-ray diffraction (XRD) can differentiate crystalline carbon-based black pigments (like graphite) from non-crystalline forms and can make further distinctions based on noncarbon components similar to XRF (*1, 2, 5*). However, it typically cannot differentiate non-crystalline, amorphous carbon-based black pigments (*2, 4*). Another study has shown good results in using powder XRD and a synchrotron beamline in identifying the type of carbon-based black pigment present in archeological samples, but required invasive sampling and powdering of the sample taken (*28*).

Nonlinear optical microscopy techniques, such as two-photon fluorescence, second-harmonic generation, and coherent anti-Stokes Raman microscopy, have been shown to provide non-invasive, high-resolution imaging contrast in applications to biology (*29*) and, more recently, cultural heritage science (*30-33*). These contrasts are easily measured as they are emissive, generating light at wavelengths different from the excitation light. However, these conventional multiphoton techniques will not aid in distinguishing carbon-based black pigments; there is little to no fluorescence to analyze (*12*), and their Ramen spectra are not pigment specific (*6, 16-19*).

We demonstrate here that another nonlinear optical technique, femtosecond pump-probe microscopy, can identify and distinguish common carbon-based black pigments non-invasively. Pump-probe microscopy takes advantage of the nonlinear interactions of two laser pulses with the sample to provide remarkable molecular specificity: in many cases there are multiple competing molecular mechanisms which provide significant contrast between nominally similar molecules. We focus here on transient absorption (TA), a subset of pump-probe, shown in figure 1, in which an excitation (pump) pulse affects the absorption of a time-delayed (probe) pulse. 'Instantaneous' mechanisms such as stimulated Raman scattering (SRS), two-photon absorption (SRS), sum-frequency generation (SFG), and cross-phase modulation (XPM) give signals only when pump and probe pulse overlap in time. Other molecular mechanisms result in delayed time signals. The pump laser pulse excites population into higher electronic states and simultaneously creates a population hole in the electronic ground state. Intermolecular vibrational redistribution rapidly rearranges the population of the electronically excited molecules, which can be transferred by the second pulse into a higher electronic state, namely excited state absorption (ESA) or to a vibrationally excited level of the ground state, namely stimulated emission (SE). The population hole in the ground state created by the pump pulse reduces the number of molecules available to be excited, reducing the absorption of the probe, a mechanism labeled ground state bleach (GSB). ESA and SE occur on roughly the same timescale, but the other effects have independent rates. Finally, pump absorption can cause localized heating, which in turn, can change the index of refraction. This affects, depending on the grain-size of the material, the Mie or Rayleigh scattering, effectively changing directional scattering, an effect called thermal scattering (TS). All of these non-emissive pump-probe interactions are separated from background signals using a modulation transfer scheme, explained in more detail in reference (*34*).

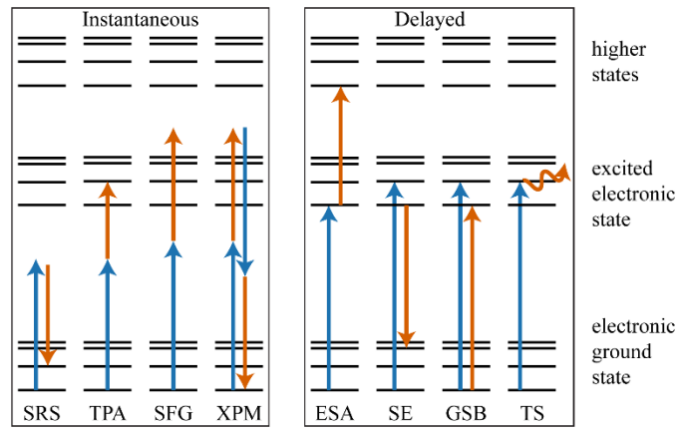

*Figure 1. Multiphoton nonlinear processes.*

Pump-probe microscopy has successfully been applied in a wide range of applications, including melanin characterization in biological tissue (*35*). This application provides a good example of the versatility: the melanin absorption spectrum is broad and featureless, but pump-probe images reveal significant heterogeneity from many competing molecular mechanisms shown in figure 1, and the contrast correlates with disease progression in melanoma. Previous cultural heritage applications include identification of iron oxides and red organic dyes, visualization of vermilion degradation, and as a tool to noninvasively obtain a virtual cross-section of historical works of art (*36-41*). Here, we use pump-probe microscopy to identify and distinguish four of the most used carbon-based black pigments, bone black, charcoal, graphite, and lamp black. We demonstrate that pump-probe microscopy reveals unique, nonlinear spectral features of these pigments that allow identification in two-compound black-black mixtures (which would be applicable in separating an underdrawing from black paint used in upper layers) and to identify black pigments in shading applications, i.e. a mixture of black with other colorful pigments, such as ultramarine blue.

**Results**

*Pump-probe spectroscopic features of black pigments*: We acquired pump-probe (P-P) image stacks which are series of P-P images at 27 different time delays ∆t with a pump wavelength of $I_{pump}$ = 720 nm and a probe wavelength of $I_{probe}$ = 817 nm. These stacks were taken from pure samples of bone black, charcoal, graphite, and lamp black. Each P-P stack consists of 128x128 pixels, and each pixel contains a P-P spectrum. These stacks of pure pigments were averaged across the spatial dimensions, that is over the imaged field of view, and normalized. The normalized spectra are shown in figure 2, while the non-normalized spectra can be found in the supplementary materials, figure S2. The P-P spectra of the four pigments exhibit distinctive qualitative differences. For graphite and lamp black, the duration of the temporal features is on the order of 100 fs, limited by the temporal resolution of our microscope. These 'instantaneous' signals suggest the involvement of virtual energy states in the nonlinear interaction, typical of processes like TPA or SRS. In our convention, transient loss processes such as TPA are depicted as positive while transient gain processes such as SRS, with the chosen pump-probe combination, are depicted as negative signals. This indicates TPA as the likely signal origin for graphite and lamp black. The spectra of bone black and charcoal are dominated by multiple excited state absorption processes, which are described by a superposition of multiple exponential decays. The obvious differences in the P-P spectra highlight the potential of P-P microscopy to noninvasively identify and distinguish these four carbon-based black pigments.

*Pure pigment heterogeneity*: The pump-probe images uncover variations within pure pigments that are not apparent in their averaged spectra. For example, the averaged P-P spectrum of bone black is uniformly positive, peaking around a time delay of Dt = 0.5 ps. However, high-resolution pump-probe images, shown in figure S3 in the supplementary material, reveal interspersed regions of positive and negative signal. A convenient way of visualizing heterogeneity in pump-probe stacks is an adapted form of phasor analysis (*42*). In phasor analysis, single-frequency sine and cosine Fourier coefficients are calculated for each pixel and plotted as the $x$- and $y$- coordinates in a phasor histogram. For example, phasor coordinates of positive (negative) single-exponential decays would map onto a specific point on the semi-circle in the first quadrant (third quadrant). Nearby points in a phasor diagram correspond to similar P-P signals. We show phasor histograms, computed with a frequency of f = 0.25 THz, of the pure black pigments in figure 3 (A, D, E, H). It is evident that the phasor histograms for bone black and charcoal fall into two distinct areas, aligning with positive and negative P-P spectra, respectively. Selecting the clusters indicated by red and yellow circles, we plot their corresponding P-P spectra in figure 3 (B, C, F, G) using the corresponding colors. For example, the phasor histogram of pure bone black is shown in figure 3A. We average all (pixel) spectra of the P-P image that fall into the colored circle in the phasor histogram and plot them in figure 3B in the corresponding color. The signals in both, bone black and charcoal, appear to be the same aside from a sign difference, respectively. This suggests TS as an underlying molecular mechanism. A pump-induced change in refractive index transiently changes the angular distribution of the backscattered light, that in combination with an aperture in the beam path causes a sign-change in the measured signal. An alternative interpretation would be the presence of two distinct chemical species in bone black and in charcoal. Investigation of the molecular origin of these signals goes beyond the scope of this manuscript as we found an effective approach to deal with this type of pigment heterogeneity. Conversely, graphite and lamp black appear homogeneous in their phasor histograms.

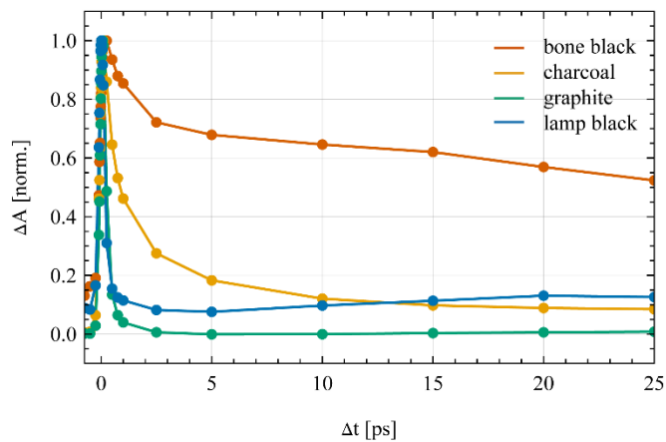

*Figure 2. Spatially averaged pump-probe spectra of bone black, charcoal, graphite, and lamp black.*

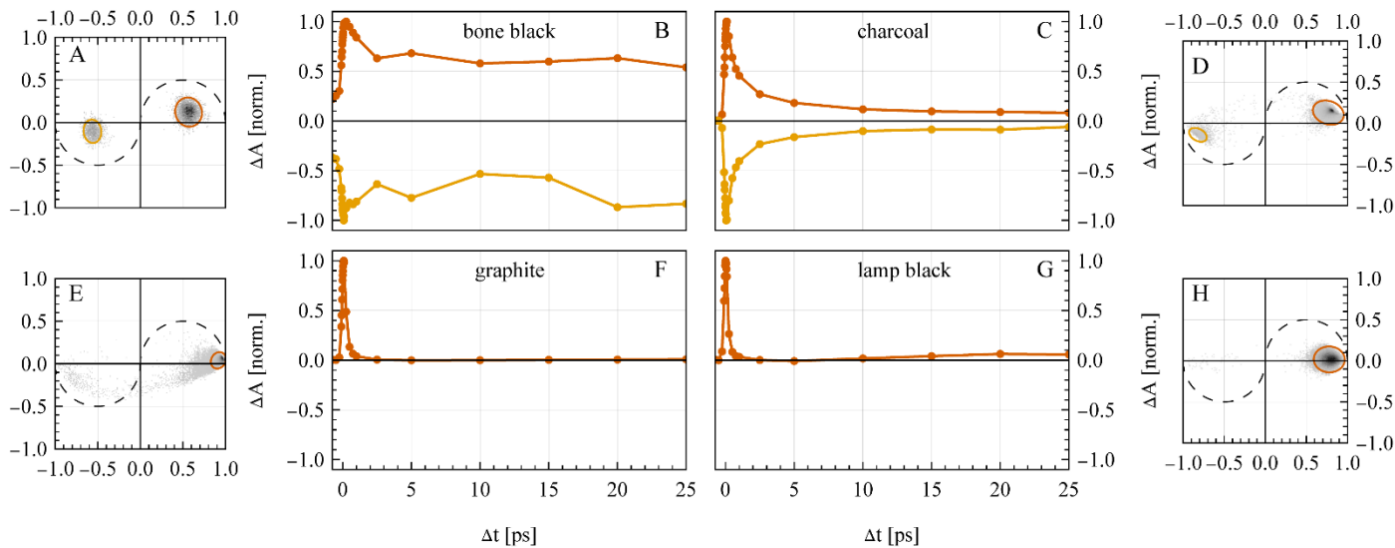

*Figure 3. Phasor histograms and pump-probe signal components of the carbon-based black pigments.* A, D, E, and H: The phasor coordinates of all signal-containing pixels in pure pigments (bone black, charcoal, graphite, and lamp black, respectively) as histograms. B, C, F, and G: Averaged P-P signals corresponding to the circled regions of the phasor histogram.

*Pigment assignment by unmixing*: While the averaged spectra of the pigments shown in figure 2 are distinctive and distinguishable, these representations obscure the underlying heterogeneity within the pigments. For a typical cultural heritage application, a specific region or volume of interest would be imaged, and we would like to derive pigment identity in this specific region. This is a common problem in many applications such as hyperspectral imaging or Raman imaging, where the spectra of reference samples are well-known. The goal is to derive an abundance map from the hyperspectral or Raman image, determining the proportion of each reference spectra within every pixel. We apply the same principle to P-P images of black pigment mixtures. We use the averaged P-P curves shown in figure 3 as reference spectra. Due to the noted heterogeneity in bone black and charcoal, phasor analysis is utilized to derive two distinct reference curves for each of these pigments. Subsequently, in the unmixing process, probabilities for the two reference spectra associated with the same pigment are combined. We used a fully constrained least square algorithm based on reference (*43*) and implemented in pysptools 0.15.0 by Christian Therien. The model incorporates a non-negativity constraint, which permits only positive coefficients in constructing the measured signal as a weighted sum of reference spectra. Additionally, it enforces the constraint that the coefficients for a single pixel sum up to one.

When presented with the black pigment mixtures, the unmixing approach correctly identified only 65% of the pixels in both the black-black mixtures and ultramarine blue-black mixtures. Given the low accuracy, we changed our approach to methods that are better suited to the complex and heterogeneous nature of the black pigment P-P data.

*Machine learning for pigment classification*: Two popular methods have been used in the past to evaluate pump-probe spectra and assign them to molecular species: Principal component analysis and model fitting. Because several nonlinear optical interactions such as TPA, SRS, and ESA contribute to the measured pump-probe signals, the resulting P-P spectra are generally bipolar superpositions of multiple exponential decays and intrinsically non-orthogonal. As principal component analysis relies on orthogonal data structures, we consider it non-ideal for identification of carbon-based black pigments. Model fitting of P-P spectra with exponential basis functions is a powerful method and could allow for pigment identification based on specific lifetimes and amplitudes. However, there exists no method to unambiguously separate the superposition of multiple exponential decays into fundamental components. Also, there are limitations on how precise amplitudes and lifetimes from exponential decays can be extracted for a given signal-to-noise level (*44, 45*). Because the pump-probe signals from black pigments are generally weak, spatial resolution would need to be sacrificed by down-sampling to achieve an appropriate signal-to-noise level for fitting.

For these reasons, and the intrinsic heterogeneity in the data, we decided to go a different route and train a support vector machine (SVM) for classification. An SVM is a supervised learning algorithm that, in its simplest form, classifies data into one of two classes. The algorithm takes n-dimensional input vectors (here pump-probe spectra consisting of 27 time delays) and separates them by a n-1 dimensional hyperplane. The hyperplane, the plane that maximizes the margin between classes, is defined by the support vectors, the data points from each class that are nearest to the hyperplane and most influence its position. An SVM can be expanded to multiclass classification with a "one-versus-rest" strategy. It naturally lends itself to heterogeneous data and is well suited for high-dimensional data. We trained a SVM with P-P spectra from pure pigments and then used it to classify and identify pigments in two-component mixtures. Opposed to the unmixing approach, we train the SVM with around 6600 spectra of each of the five pigments (bone black, lamp black, charcoal, graphite, and ultramarine blue), thereby exposing the SVM to the full range of pigment heterogeneity. A description of the training, validation, and testing process can be found in the materials and methods section of this manuscript.

In brief, the resulting SVM has an overall accuracy of 96% for pure pigments. This means that when the SVM is presented with a P-P spectrum from a single pixel of any of the five pigments (bone black, lamp black, charcoal, graphite, and ultramarine blue), it classifies it correctly 96 times out of 100. However, the performance of the SVM will drop for two-pigment mixtures to around 80%, as discussed in the next sections.

*Black-Black Mixtures*: The accuracy reported above was obtained when identifying pure pigment samples. To investigate the SVM performance in a more relevant application, we tested it on six 50-50 (by weight) pigment mixtures of two different black pigments: bone black-charcoal, bone black-graphite, bone black-lamp black, charcoal-graphite, charcoal-lamp black, and graphite-lamp black. We took images of three different areas for each mixture and presented the P-P image stacks to the SVM classifier trained on pure pigments.

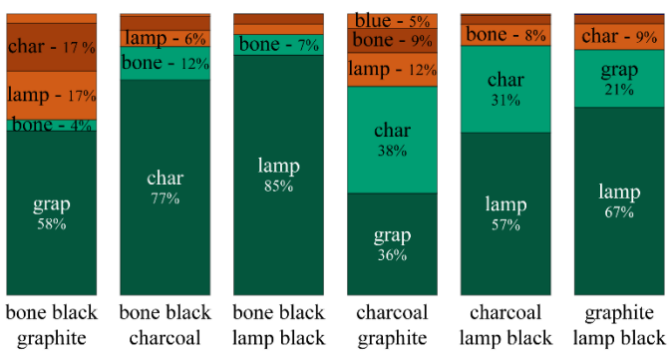

*Figure 4. Summary of SVM performance on two-black pigment mixtures.* The bar charts display the breakdown for classification of each black-black mixture. Each bar adds up to 100%. Green corresponds to correctly classified pixels. Red corresponds to misclassified pixels.

We summarize the overall performance of the classifier with a bar chart in figure 4 one bar for each black-black mixture. It is crucial to note that we operate without a definitive "ground truth" in this context. While we know the two pigments comprising each mixture, their precise microscopic distribution remains unknown, and no alternative method exists for validating the P-P microscopy and classification results. To assess performance, we tally all classified pigments for a given mixture. We label pixels as correctly classified if they are part of the mixture and as misclassified if identified pigments are not part of the mixture. A complete table can be found in the supplementary information, table S1. A successful method to identify black pigments should have a high accuracy, meaning that the percentages of correctly identified pixels should be larger than any other misclassification. In our case the accuracy ranges between 62% (bone black-graphite) and 93% (bone black-lamp black). We will discuss the implications of these numbers in the discussion section.

The SVM returns a pigment classification for each pixel in the P-P stack, and we use this information to generate a false-color pigment map (or abundance map). Three representative examples are shown in figure 5. For the charcoal-graphite mixture in panel A, 71% of the signal containing pixels are identified as charcoal (39%) or graphite (32%). The remaining 29 % were misclassified, the worst offender being lamp black with 16 % of the pixels. The graphite-lamp black mixture, shown in figure 5B, performs well with 87% of the signal containing pixels being correctly identified (25% graphite, 62% lamp black). The largest misidentification is charcoal (11%). In the bone black-charcoal mixture, figure 5C, 87% of pixels are accurately identified and the largest offenders are graphite and lamp black, each with 7% of the pixels. While these samples were prepared as 50-50 mixtures by mass, each microscopic region in figure 5 deviates from that distribution. We do not expect to precisely measure the preparation ratio in a microscopic image of 36μm x 36μm.

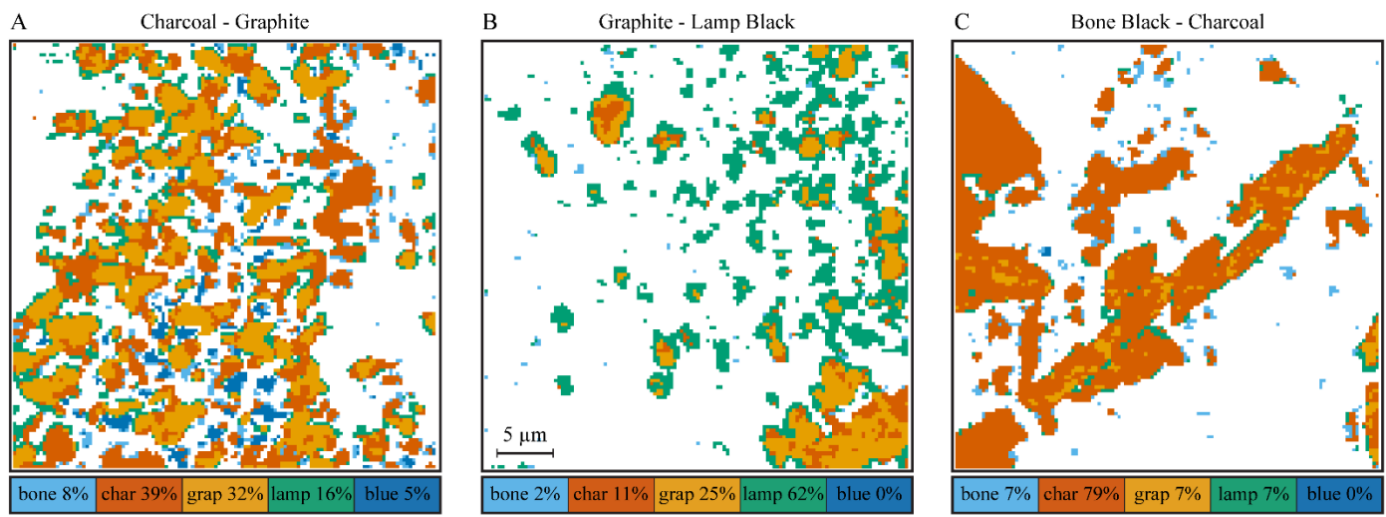

*Figure 5. Pigment map for three black-black mixtures.* A: charcoal-graphite. B: graphite-lamp black. C: bone black-charcoal.

*Ultramarine Blue-Black Mixtures*: In order to demonstrate P-P classification for shading applications, we imaged mixtures of each carbon-based black pigment with ultramarine blue ($Na_7Al_6Si_6O_{24}S_3$), a modern equivalent to its natural form, lapis lazuli. P-P information for ultramarine blue is shown in the supplementary materials, figure S4. We captured images of twelve combinations, mixing each of the four black pigments with ultramarine blue in three mass ratios: 25:75, 50:50, and 75:25 percent. Brightfield images are in figure S5. For each of the twelve samples we imaged three different areas and computed the average classification accuracy. The effectiveness of the classifier across all the shading combinations is presented in figure 6, see table S2 with full details. Like in the black-black mixture case, there exists no ground-truth and we therefore apply the same metric to measure performance. Our methodology yields a robust qualitative classification, demonstrating a strong correlation between the detected amount of ultramarine blue and the actual physical mixing ratio of the sample. This correlation is consistent for most of the black pigments used. The classification accuracy ranges between 59% for the 75:25 lamp black-ultramarine blue mixture and 96% for the 50:50 bone black-ultramarine blue and 25:75 lamp black-ultramarine blue mixtures. Besides the two lowest performing instances (59% and 62%), the remaining 10 scenarios classify more than 70% correctly with a mean of 86%. The discussion section will further explore the probable causes for this range in performance and suggest improvements.

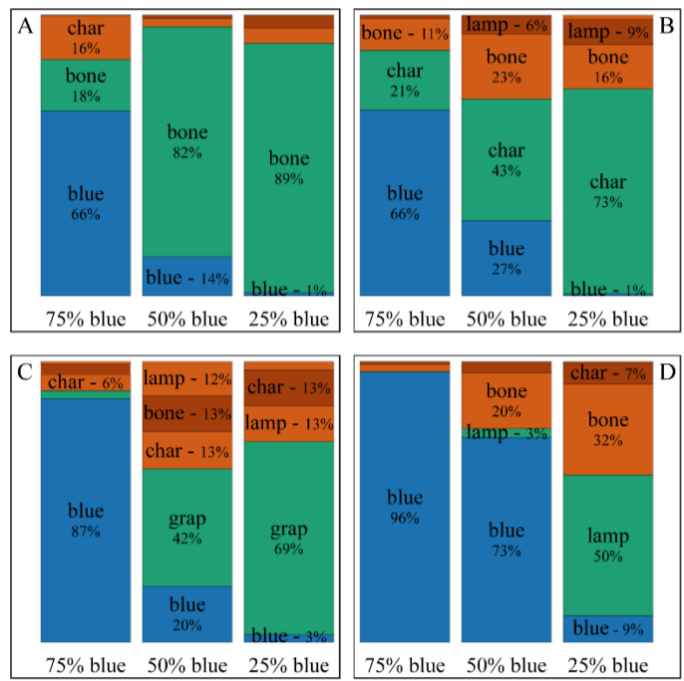

*Figure 6. Summary of SVM performance on blue - black mixtures.* A: bone black-ultramarine blue. B: charcoal-ultramarine blue. C: graphite-ultramarine blue. D: lamp black-ultramarine blue. Each bar adds up to 100%. Blue and green correspond to correctly classified pixels, blue for ultramarine blue and green for the black pigment. Red corresponds to the misclassified pixels.

Analogous to the black-black mixtures, we can derive spatial maps of the black-blue mixtures. Three example maps of the mixtures, 75:25 ultramarine blue-charcoal, 75:25 ultramarine blue-graphite, and 50:50 ultramarine blue-bone black, are shown in figure 7. In all parts, the derived mixing ratios deviate from the real, macroscopic, mixing ratio. This highlights again the challenge in the absence of ground truth and the microscopic variation of mixing ratios. However, the images shown in figure 7, as well as the bar chart in figure 6, demonstrate good qualitative performance.

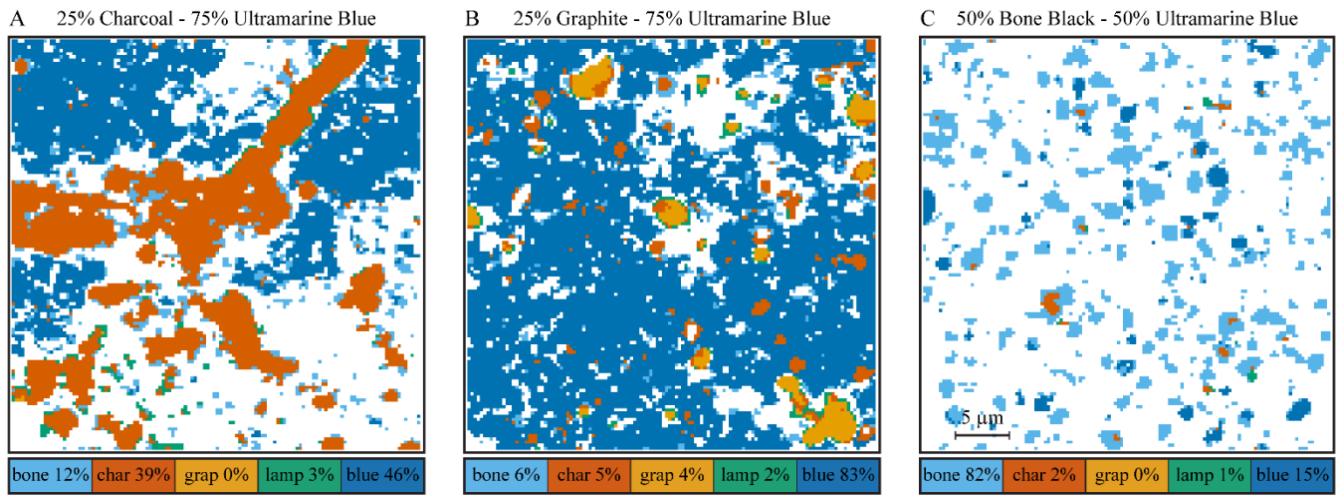

*Figure 7. Pigment map for three ultramarine blue-black mixtures.* A: 75:25 ultramarine blue-charcoal. B: 75:25 ultramarine blue-graphite. C: 50:50 ultramarine blue-bone black.

## Discussion

Our findings highlight the potential of pump-probe microscopy for the noninvasive differentiation and mapping of black pigments. This technology presents a significant advancement in the noninvasive analysis of cultural heritage artifacts, where the precise identification of pigments can provide invaluable insights into the techniques and materials used by artists. However, there are cases of misclassification within the set of targets. Here we will discuss the performance and limitations of our current classification approach and then comment on the future application of pump-probe microscopy to more practical scenarios.

*Classification challenges and future strategies*: Our current classification approach faces several challenges. The primary challenge we encounter with our current classification approach is the absence of a definitive ground truth for the images. The derived pigments maps, as shown in figure 5, are a result of P-P imaging and a SVM classifier. However, we know of no alternative methods to validate the correct *mapping*. The only ground truth we have is the mixing ratio used during sample preparation. The mixing ratio is a macroscopic quantity and the P-P images in this proof-of-principle study only sample three 36μm x 36μm areas. Within this small area, we expect variations in the pigment ratios that will vary from the macroscopic pigment distribution. For example, the measured pigment ratio in the 50:50 graphite-lamp black mixture, in figure 5, is 25/62 ≈ 0.4 which could be explained by a locally higher density of lamp black, rather than an inaccuracy in our method.

A second challenge in pigment classification is the vast differences in signal-to-noise ratio (SNR) of different pigments. We show unnormalized curves of the four black pigments in the supplementary materials, figure S2. For example, typical bone black P-P signals are around 50 times weaker, and typical lamp black signals around 5 times weaker than signals of either charcoal or graphite. Signals of bone black and lamp black are close to the noise floor of our current microscope and are therefore more difficult to classify compared to the high SNR signals of charcoal and graphite. We believe that this is the main reason for the small bone black percentages identified in black-black mixtures, such as in figure 4. This also relates to the inherent tradeoff between spatial resolution and SNR: higher spatial resolution requires smaller sampling volumes and therefore leads to smaller SNR. Averaging neighboring pixels in an image increases SNR at the expense of spatial resolution. Spatial averaging, however, might mix spectra of adjacent pigments. This causes two additional challenges. In case of a large signal amplitude mismatch, the pigment with the larger signal will bury the spectral features of the weaker pigment signal thereby skewing classification towards pigments with larger signal. If spectra of comparable signal strengths are averaged, we create a de facto new spectrum that is unknown to the SVM classifier, which was only trained on pure spectra, leading to increased misclassification. Currently, we choose a compromise between spatial averaging to achieve sufficient SNR and maintaining enough resolution to resolve most individual pigment grains. In the future, we envision increasing SNR in two ways. First, by improving our detection capabilities, specifically by increasing collection efficiency by using higher numerical aperture objectives. Second, we intend to explore different pump and probe wavelengths that might offer a larger interaction cross-section for the currently weak signals of bone and lamp black.

We also observe, at least heuristically, an increased misclassification rate at pigment grain boundaries. This can be seen in figure 7A and B, in which bone black appears around the edges of individual pigment grains. This can also be observed for other pigments such as lamp black. Whether this is caused by the grain boundary itself, by data thresholding, by the intrinsically weak SNR of bone and lamp black, or by the choice of classifier algorithm needs to be investigated in the future.

An additional challenge is the intrinsic heterogeneity of pump-probe signals in black pigments. We hypothesize that two factors contribute to this heterogeneity: First, the inherent microscopic heterogeneity of charcoal and bone black. Of the four pigments, only graphite has an ordered molecular structure (sheets of sp2 hybridized carbon), consistent with the observed homogeneous signal in the phasor histogram in figure 3E. Lamp black, which undergoes a gas phase carbonization during production, is microscopically uniform (on the scale of the resolution of our microscope). Charcoal and bone black are derived from extremely heterogeneous base material, e.g. wood and bone, respectively, maintain a solid or liquid structure during carbonization, and therefore retain part of their initial structural complexity. The other factor that could contribute to the heterogeneity in pump-probe signal is the presence of heteroatoms or non-carbon constituents that commonly occur in carbon-based black pigments. Winter reports that incorporation of heteroatoms into the carbon matrix during the carbonization process is especially common for cokes and chars prepared at low temperatures (*2*). This may support the higher degree of heterogeneity of signals in charcoal and bone black. Heteroatoms can also be introduced via non-carbonaceous compounds that carry over from the source material. For example, bone black also contains the inorganic material hydroxyapatite, which would further explain why bone black displays more heterogeneity than charcoal.

In our current classification scheme, we do not utilize local image information for classification. As of now, each pixel is classified independently from each other. However, given a specific pixel, there is a high probability that adjacent pixels are of the same pigments because pigments grains extend over multiple pixels. A logical next step is to explore classification with convolutional neural networks that are designed to respect local image information.

*Beyond proof-of-principle studies towards applications to works of art*: This manuscript demonstrates the potential of P-P microscopy to noninvasively identify black pigments in mixtures. For this proof-of-principle demonstration we restricted ourselves to four black pigments and one colored pigment. Most works of art contain many more colors and, although we used the four most prevalent black pigments, there are other black pigments in use. Our group has analyzed a range of pigments, including red organic dyes, iron oxides, vermillion, and cadmium sulfide and we can add these pigments to our classification scheme (*36-41*). Furthermore, P-P microscopy offers two powerful and easily accessible degrees of freedom: the choice of pump and probe wavelength. Pump-probe signals reflect the population dynamics between molecular levels and are therefore strongly dependent on the pump and probe wavelengths. Pigments that present similar spectra at a particular wavelength combination may differ drastically at another (*37*). A convenient approach would be pigment exploration in a broadband pump-probe spectroscopy setup, where many wavelengths can be probed simultaneously. Then we would select a wavelength combination that offers a unique contrast for a specific pigment. Ultimately, multiple P-P images acquired with different wavelength combinations (hyperspectral pump-probe microscopy) will provide sufficient specificity to distinguish and identify many pigments. Extension of the SVM classifier to more pigments and to hyperspectral P-P images is conceptually straight forward and only requires the additional pure reference data in the training phase. In addition, polarization pump-probe microscopy, which offers improved chemically specific contrast based on the molecular anisotropy of pigments, can further improve pigment specificity (*35, 46*).

The multiphoton nature of pump-probe microscopy enables high resolution in all three spatial dimensions, even beneath the surface of highly scattering materials (*29*). In previous experiments, we were able to image up to a depth of $\approx 90~\mu m$ in paint layers to produce virtual cross sections (*39*), thus allowing cultural heritage scientists and conservators to better understand pieces of art without invasive sampling. However, achievable penetration depths depend on the absorption and scattering properties of the materials present at the surface and the subsequent layers. Carbon-based black pigments strongly absorb visible to near-infrared light and therefore reduce optical penetration depth. This will be most prominent in works of art with a thick or opaque layer of carbon-based black paint, for example in oil or tempera paintings. In a work with thinner or more transparent layers, such as watercolor paintings or drawings and prints, the absorption of the black pigments would not significantly reduce penetration depth.

Our study has successfully shown that pump-probe microscopy is an effective noninvasive tool for differentiating black pigments in a variety of combinations, including mixtures with other blacks and with ultramarine blue. Across 18 different blends and shades, the technique accurately identifies 80% of the pixels in each image. This achievement highlights pump-probe microscopy's capability to fill a significant void in the field of cultural heritage science, where, until now, no noninvasive method for identifying black pigments existed. We have outlined a clear strategy to further improve the performance and to increase the number of pigments in our approach and we envision applying this methodology to actual works of art. A particularly fascinating application would be Vermeer's Girl with a Pearl Earring where bone black and charcoal are reported to exist together in an underlayer, currently only confirmed by analysis of a cross section (46). Pump-probe microscopy could be used to further validate these findings as well as to provide new information, i.e. a three-dimensional pigment map of both pigments across the painting.

## Materials and Methods

*Pump-Probe Microscopy*: A schematic of our pump-probe microscope is shown in figure S1. The output of a Ti:Sapphire laser (Coherent Chameleon II) with an 80 MHz repetition rate is split into two parts. One part serves as probe beam at a wavelength of $\lambda_{probe}$ = 817 nm. The second part is frequency converted into the pump with a wavelength of $\lambda_{pump}$ = 720 nm with an optical parametric oscillator (Coherent Mira-OPO). The pump pulse train is intensity-modulated by an acousto-optical modulator at a rate of 2 MHz. Both laser beams are spatially superimposed, sent into a laser scanning microscope, and focused onto the sample with a 20x 0.7 NA dry objective. The inter-pulse delay ∆t between pump and probe is controlled with a translation stage in the probe beam path. We utilize a modulation transfer scheme to detect the weak signals generated by the nonlinear interaction between pump, probe, and sample. As the nonlinear interaction transfers the pump modulation onto the probe pulse train, these changes in absorption in the probe pulse train are measured with a photodiode and a lock-in amplifier. For pigment imaging we use a pump and probe pulse intensity of $I$ = 4.4x10$^8$ $W/m^2$, (corresponding to 0.25 mW), and image an area of 36µm x 36µm for 27 time delays ∆t spanning -10 to 25 ps. Thus, the resulting data structure (image stack) is a 3-dimensional data cube with two spatial and one temporal dimension. Each pixel in the pump-probe stack represents a P-P spectrum, the change of absorption as a function inter-pulse delay ∆t.

*Preparation of Pigments*: Pigments were thoroughly mixed with gum Arabic in a separate vessel to prepare a smooth, watercolor paint. This paint was applied to a commercially prepared canvas in two layers, allowing for drying in between. The pigments were commercially sourced from AGS Company (graphite), Coates Charcoal (charcoal), and Rublev Colours (bone black and lamp black).

*Data pre-processing of P-P image stacks*: Raw P-P data is pre-processed before training and classification in the following steps: (1) Due to pump-leakage into the probe beam and potential long-lived ($\tau \gg$ 12.5 ns, the time spacing between consecutive pulses) radiative states at the probe wavelength, we average three P-P images at negative time delays (∆t = -10 ps, -5 ps, and -2.5ps) and subtract them from the entire P-P stack, thereby eliminating a constant offset in the data. (2) To increase SNR, we apply a spatial moving average filter of kernel size two. (3) A global intensity threshold is applied to all P-P stacks to discriminate noise from actual P-P signals. (4) We reduce the image size by down sampling of a factor two, consistent with the moving average filter. (5) For pure pigment images we apply their corresponding phasor masks, shown in Figure 2. For pigment mixtures, we apply a phasor mask that includes a combination of phasor masks from all individual pigments.

*Support Vector Machine (SVM)*: We use the scikit-learn 1.4 (*47*) and the imbalanced-learn 0.12.0 (*48*) python packages for training, validation and testing of the SVM. We randomly select around 6600 pure pigment, single pixel P-P spectra (imblearn RandomUnderSampler) for ultramarine blue, bone black, lamp black, charcoal, and graphite. We split them into training and testing set with a ratio of 3:1. We perform a hyperparameter optimization (scikit-learn GridSearchCV) of $C$ and $\gamma$ for the SVM (scikit-learn SVC) with the training set, a stratified-5-fold strategy (scikit-learn StratifiedKFold), accuracy as scoring metric, and a standard scaler applied to all spectra (scikit-learn StandardScaler). The performance of the best classifier is inferred by measuring accuracy of the classifier applied on the test set.

The entire procedure is repeated 5 times, and we compute the average and standard deviation over all five runs. The validation accuracy of $acc_{valid}$ = (96.32 ± 0.08)% and the testing accuracy of $acc_{test}$ = (96.47 ± 0.25)% are comparable and let us conclude that the classifier is well-trained. We then use the best hyperparameters and the entire data set from homogeneous pigment samples (5 x 6600 spectra) to train the final SVM classifier that is used to classify in two-pigment mixtures. Note that the final classifier is solely trained on homogeneous pigment data and has not been trained with any data from mixed samples.

## Acknowledgments
We thank Yuxiao (Michelle) Wei for stimulating discussions and support for spectral unmixing. We thank Yue Zhou and Jacob Lindale for discussions and support regarding pump-probe microscopy and cultural heritage science.

**Funding:** This material is based upon work supported by the National Science Foundation Division of Chemistry under Award No. CHE-2108623 (M.C.F.). Martin Fischer also received salary support from the Chan Zuckerberg Initiative (grant no. 2021-242921).

**Author contributions:** Conceptualization and Methodology: HVK, DG, MCF, WSW; Investigation: HVK, MT; Formal Analysis: DG; Visualization: HVK, DG, MCF; Writing-original draft: HVK, DG; Writing-review & editing: HVK, DG, MCF, WSW, MT; Funding acquisition and Supervision: WSW, MCF.

**Competing interests:** Authors declare that they have no competing interests.

**Data and materials availability:** All data needed to evaluate the conclusions in the paper are present in the paper and/or the Supplementary Materials. Pump-probe data sets will be made available on the Duke University Research Data Repository at DOI:10.7924/r43j3nx2k . Additional data related to this paper may be requested from the authors.

**Supplementary Materials**

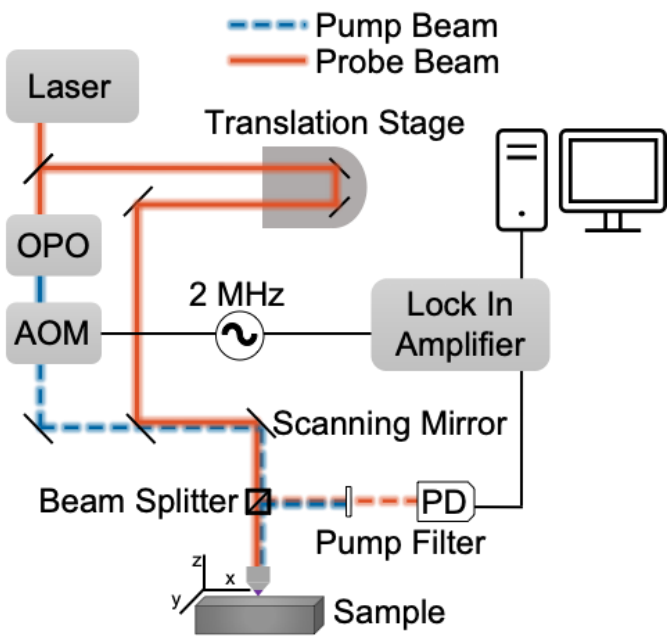

**Figure S1.**
**Schematic experimental set-up for pump-probe microscopy.** A femtosecond laser at $\lambda_{probe} = 817$nm is split into two spatial modes. One mode serves as pump for an optical parametric oscillator (OPO) creating the pump wavelength at $\lambda_{pump} = 720$nm, which is also amplitude modulated at a frequency of 2 MHz with an acousto-optic modulator (AOM). Pump and probe beam are spatially overlapped and coupled into a conventional laser scanning microscope.

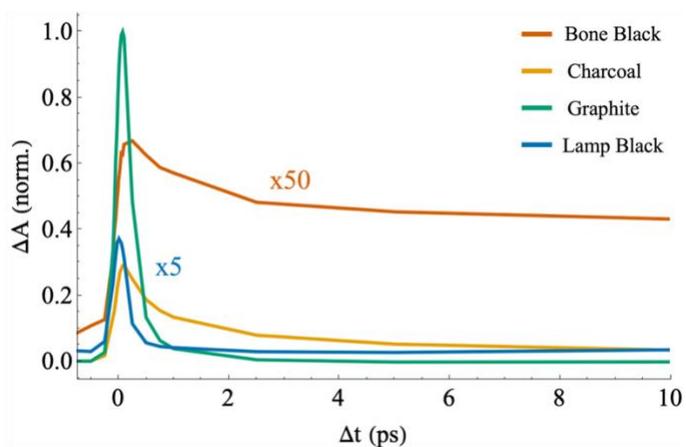

**Figure S2.**
**Pump-probe spectra of bone black, charcoal, graphite, and lamp black.** Each curve was normalized to the Graphite maximum, and then adjusted (bone black x50, charcoal x1, graphite x1, lamp black x5) for visual comparison.

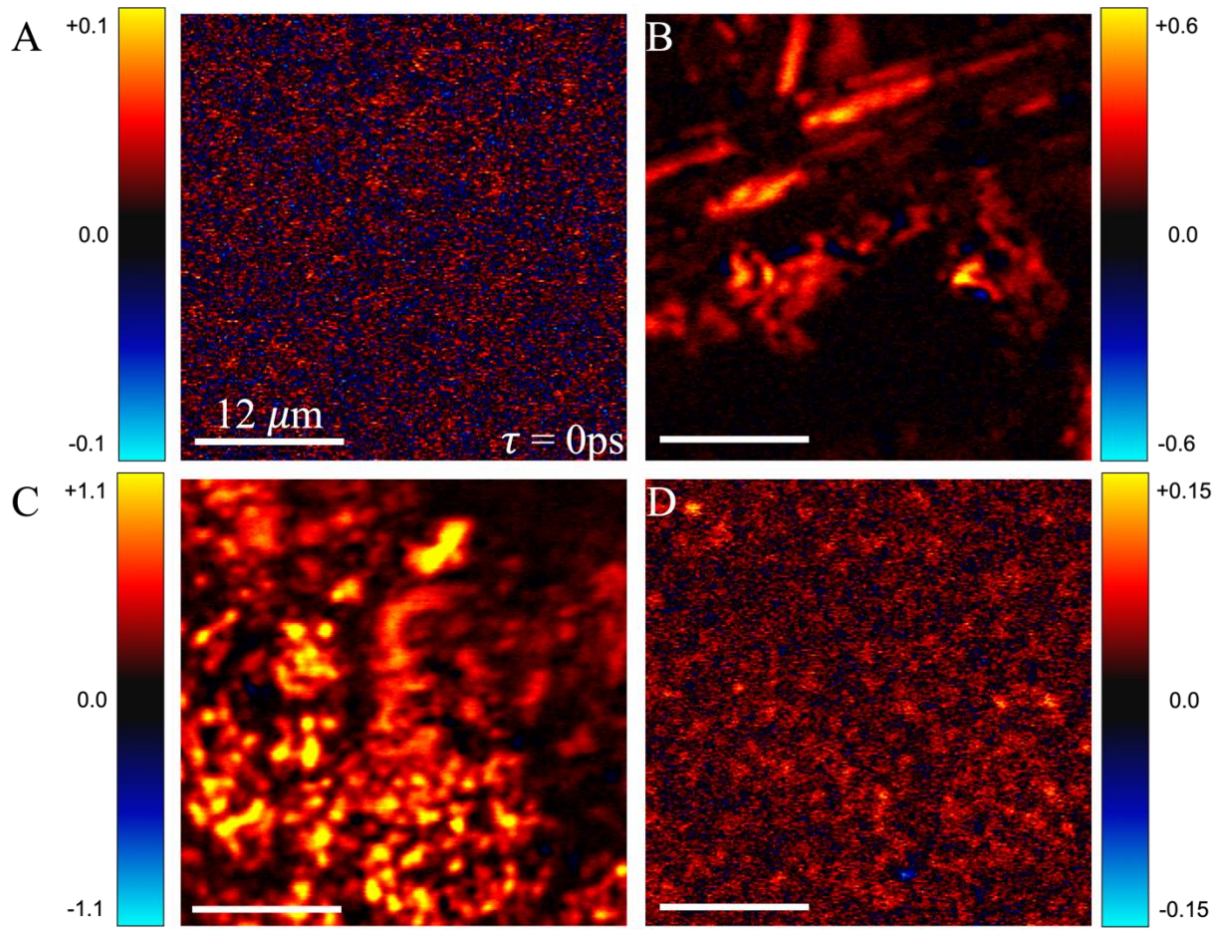

**Figure S3.**

**Pump-probe images for bone black, charcoal, graphite, and lamp black.** Each image represents the pump-probe image at a time delay of $\Delta t=0$ps of an area of 36μm x 36μm for A) Bone Black B) Charcoal C) Graphite D) Lamp Black

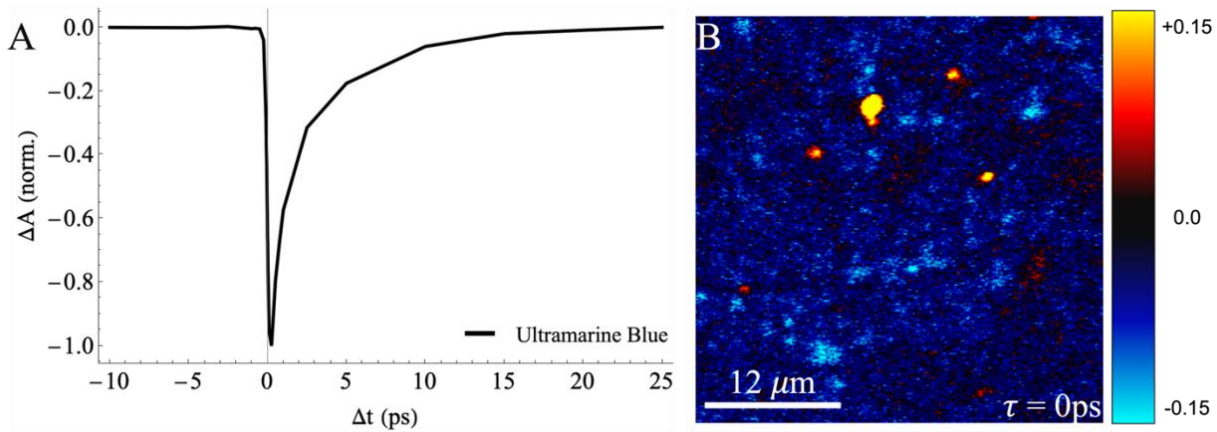

**Figure S4.**

**Average pump-probe spectrum and pump-probe image for ultramarine blue.** An average pump-probe spectra of ultramarine blue, corresponding the region of interest (36 μm x 36 μm) shown in part B at a time overlap $\Delta t=0$ps.

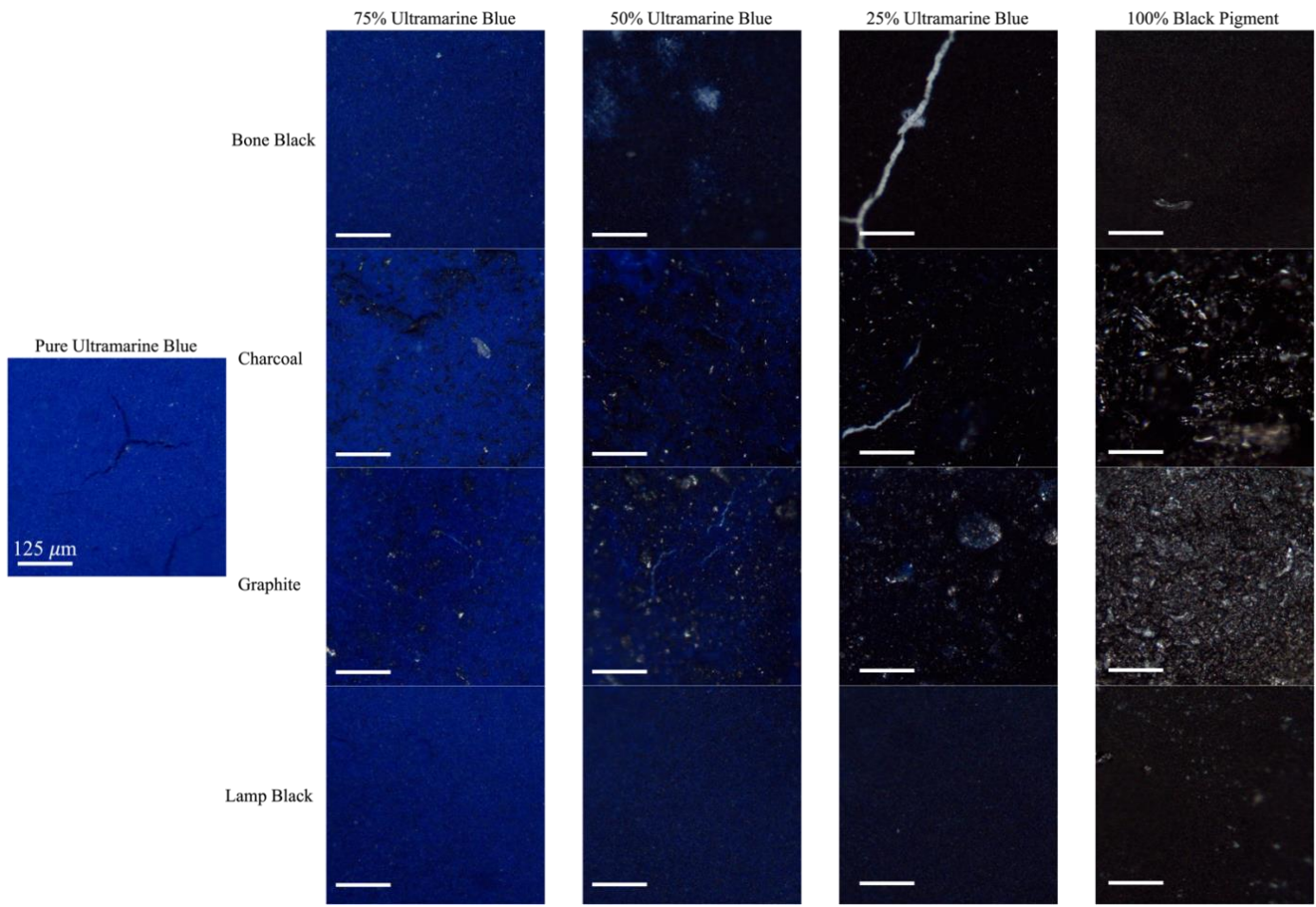

**Figure S5.**
**Brightfield images of the twelve shading mixtures and the five pure pigments.** The standalone leftmost image is of pure ultramarine blue. The columns going from left to right the images are decreasing amounts of ultramarine blue, 75%, 50%, 25%, 0%, with the last column being pure black pigment. The rows, from top to bottom, are bone black, charcoal, graphite, and lamp black. Each image is a 500 x 500 micrometer field of view, the scale bar represents 125 micrometers.

| Black-Black Mixture | Bone Black [%] | Charcoal [%] | Graphite [%] | Lamp Black [%] | Ultramarine Blue [%] |
|---|---|---|---|---|---|
| Bone – Char | 12 | 76 | 5 | 6 | 1 |
| Bone – Grap | 4 | 17 | 58 | 17 | 4 |
| Bone – Lamp | 7 | 4 | 4 | 85 | 0 |
| Char – Grap | 9 | 38 | 36 | 12 | 5 |
| Char – Lamp | 8 | 31 | 3 | 57 | 1 |
| Grap – Lamp | 2 | 9 | 21 | 67 | 1 |

**Table S1.**
**Support Vector Machine classification accuracy for black-black mixtures.** Abbreviations: Bone: Bone Black; Char: Charcoal; Grap: Graphite; Lamp: Lamp Black.

| Shading Mixture | Bone Black [%] | Charcoal [%] | Graphite [%] | Lamp Black [%] | Ultramarine Blue [%] |
|---|---|---|---|---|---|
| 75% Blue – 25% Bone | 18 | 16 | 0 | 0 | 66 |
| 50% Blue – 50% Bone | 82 | 3 | 0 | 1 | 14 |
| 25% Blue – 75% Bone | 89 | 5. | 0 | 5 | 1 |
| 75% Blue – 25% Char | 11 | 21 | 0 | 2 | 66 |
| 50% Blue – 50% Char | 23 | 43 | 1 | 6 | 27 |
| 25% Blue – 75% Char | 16 | 73 | 1 | 9 | 1 |
| 75% Blue – 25% Grap | 4 | 6 | 2 | 1 | 87 |
| 50% Blue – 50% Grap | 13 | 13 | 42 | 12 | 20 |
| 25% Blue – 75% Grap | 2 | 13 | 69 | 13 | 3 |
| 75% Blue – 25% Lamp | 3 | 1 | 0 | 0 | 96 |
| 50% Blue – 50% Lamp | 20 | 4 | 0 | 3 | 73 |
| 25% Blue – 75% Lamp | 32 | 7 | 2 | 50 | 9 |

**Table S2.**

**Support Vector Machine classification accuracy for shading mixtures.** Abbreviations: Blue: Ultramarine Blue; Bone: Bone Black; Char: Charcoal; Grap: Graphite; Lamp: Lamp Black.